\documentclass[aps,amsmath,amssymb,prl,reprint,superscriptaddress]{revtex4-1}

\usepackage{array}
\usepackage{amsmath}
\usepackage{amsgen}
\usepackage{amsfonts}
\usepackage{amsbsy}
\usepackage{amssymb}

\usepackage{braket}
\usepackage{color}

\usepackage{graphicx}
\usepackage{dcolumn}
\usepackage{bm,textcomp}

\usepackage[figuresleft]{rotating}
\usepackage{units}

\def\idty{{\leavevmode\rm 1\mkern -5.4mu I}} 

\def\Cx{{\mathbb C}}     
\def\Ir{{\mathbb Z}}     

\def\tr{\mathop{\rm Tr}\nolimits}

\mathchardef\ree="023C \mathchardef\imm="023D  

\def\HH{{\mathcal H}}

\newcommand{\beq}{\begin{equation}}
\newcommand{\eeq}{\end{equation}}

\newcommand{\rem}[1]{}

\newcommand{\al}{\alpha}
\newcommand{\G}{\Gamma}

\begin{document}
\title{Quantum Walks with Non-Orthogonal Position States}

\author{R. Matjeschk}
\email{robert.matjeschk@itp.uni-hannover.de}
\affiliation{Institut f\"ur Theoretische Physik, Leibniz Universit\"at Hannover, Appelstr. 2, 30167 Hannover, Germany}
\author{A. Ahlbrecht}
\affiliation{Institut f\"ur Theoretische Physik, Leibniz Universit\"at Hannover, Appelstr. 2, 30167 Hannover, Germany}
\author{M. Enderlein}
\affiliation{Albert-Ludwigs-Universit\"at Freiburg, Physikalisches Institut, Hermann-Herder-Str. 3, 79104 Freiburg, Germany}
\author{Ch. Cedzich}
\affiliation{Institut f\"ur Theoretische Physik, Leibniz Universit\"at Hannover, Appelstr. 2, 30167 Hannover, Germany}
\author{A. H. Werner}
\affiliation{Institut f\"ur Theoretische Physik, Leibniz Universit\"at Hannover, Appelstr. 2, 30167 Hannover, Germany}
\author{M. Keyl}
\affiliation{ISI Foundation, Via Alassio 11/c, 10126 Torino - Italy}
\author{T. Schaetz}
\affiliation{Albert-Ludwigs-Universit\"at Freiburg, Physikalisches Institut, Hermann-Herder-Str. 3, 79104 Freiburg, Germany}
\author{R. F. Werner}
\affiliation{Institut f\"ur Theoretische Physik, Leibniz Universit\"at Hannover, Appelstr. 2, 30167 Hannover, Germany}
\date{\today}

\begin{abstract}
Quantum walks have by now been realized in a large variety of different physical settings. In some of these, particularly with trapped ions, the walk is implemented in phase space, where the corresponding position states are not orthogonal. We develop a general description of such a quantum walk and show how to map it into a standard one with orthogonal states, thereby making available all the tools developed for the latter. This enables a variety of experiments, which can be implemented with smaller step sizes and more steps. Tuning the non-orthogonality allows for an easy preparation of extended states such as momentum eigenstates, which travel at a well-defined speed with low dispersion. We introduce a method to adjust their velocity by momentum shifts, which allows to experimentally probe the dispersion relation, providing a benchmarking tool for the quantum walk, and to investigate intriguing effects such as the analog of Bloch oscillations.
\end{abstract}
\maketitle

Quantum Walks (QWs) are a widely used model system for transport processes. Initially introduced from a computer science perspective \cite{Kempe2003, Grover2001, Ambainis2007, Shenvi2003, Shikano2010}, the field has significantly expanded and is now largely treated from a physics perspective \cite{Engel2007, Mohseni2008, Plenio2008, Oka2005, Ahlbrecht2011a}. In fact, ``quantum walk'' is now widely taken to be synonymous with ``discrete time/ discrete space quantum dynamics'' of a particle with internal degrees of freedom. On one-dimensional lattices, a QW can always be implemented by a concatenation of coin operations and successive state-dependent shifts  \cite{Nguyen}. Already these one-body systems are capable of simulating various physical effects such as Anderson localization \cite{Ahlbrecht2011b} or the formation of molecules \cite{Ahlbrecht2011a}. In particular, single-particle QWs are a basic building block in a bottom-up approach towards general-purpose multi-particle simulation environments \cite{Gross2009}. Therefore, one of the main interests in QWs is the possibility to study key features of quantum dynamics in a setting which can be controlled experimentally with high precision.

Experimentally, QWs have been implemented  in several different ways, for example using nuclear magnetic resonance \cite{Ryan2005}, atoms in optical lattices \cite{Karski2009}, trapped ions \cite{Schmitz2009a, Matjeschk2012, Zaehringer2010} or photonic systems \footnote{The photonic experiments model continuous QWs, which are not considered here.} \cite{Bouwmeester1999, Schreiber2010, Broome2010, Perets2008, Peruzzo2010, Schaetz2011, Owens2011, Sansoni2012}.

\begin{figure}
\includegraphics[width=8.6cm]{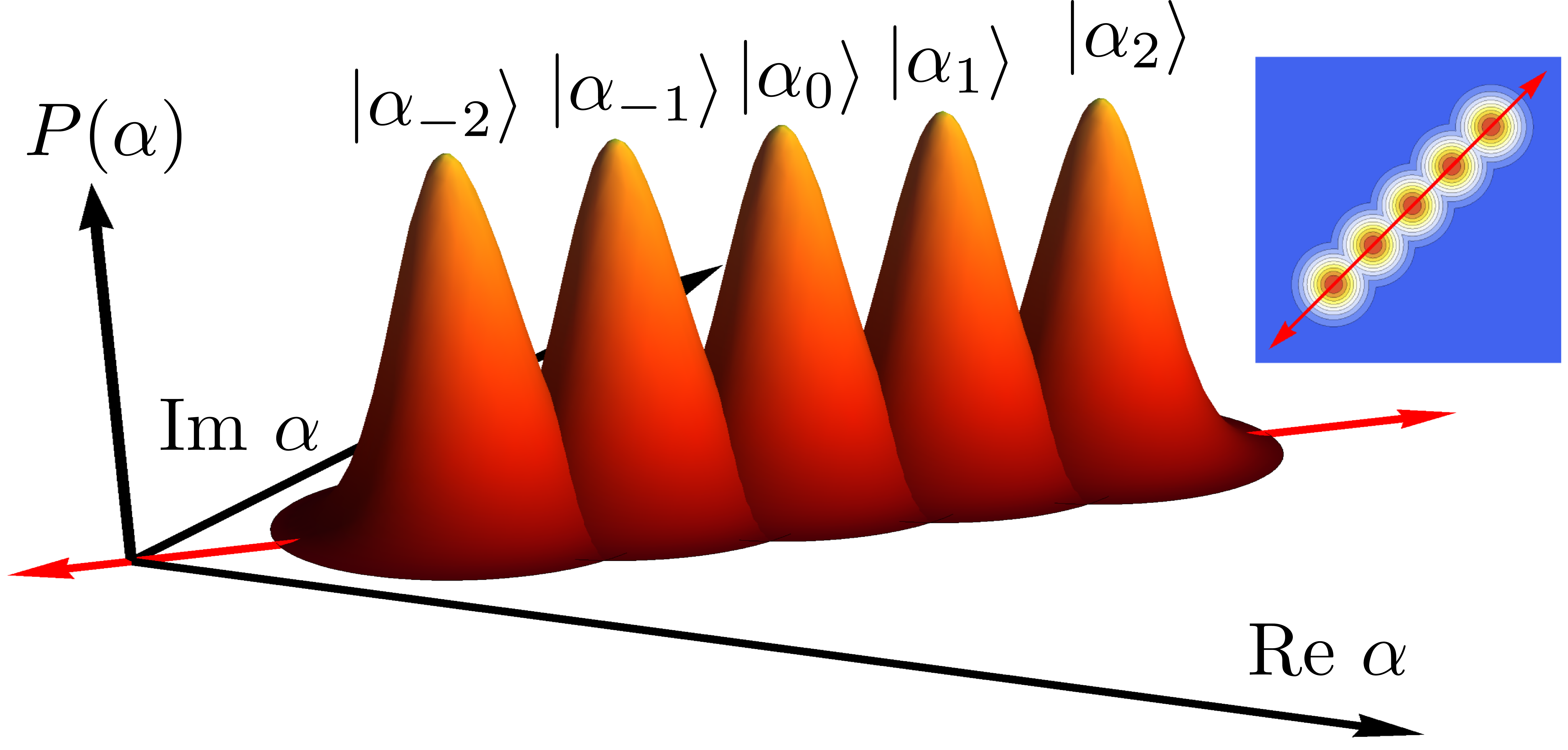}
\caption{(Color online) nQW implemented in a harmonic-oscillator phase space. The positions are coherent states, illustrated by their Husimi functions, $P(\alpha)=\lvert\braket{\alpha|\alpha_x}\rvert^2$, for each $\ket{\alpha_x}$ separately. The inlay illustrates their orientation in phase space, implementing a nQW along a line. Since the position states $\ket{\alpha_x}$ are coherent states, they are not orthogonal. The step size $\Delta\alpha=\lvert \alpha_{1}-\alpha_0\rvert=\sqrt{2}/\sigma$ (Cf. eq. \eqref{EQ:overlap}) of the nQW determines their overlap. QWs of this type have been implemented with trapped ions \cite{Schmitz2009a, Matjeschk2012, Zaehringer2010}.
}
  \label{Fig:zero}
\end{figure}

In the theoretical description it is almost universal practice, to model the different ``positions'' by mutually orthogonal subspaces in Hilbert space. However, orthogonality cannot be achieved in some proposals \cite{Travaglione2002,Xue2009,Sanders2003} and the related  experiments \cite{Schmitz2009a, Matjeschk2012,Zaehringer2010}, which use coherent states of a harmonic oscillator for the position states (Fig.~\ref{Fig:zero}). In order to fit the theoretical model, it was necessary in the experiment to choose the step size in phase space sufficiently large to make these states approximately orthogonal.

The aim of this letter is to show that the lack of orthogonality can be exploited. First of all, we give a complete analysis of QWs with non-orthogonal position states (nQWs), and introduce a transformation to the orthogonal case, such that all results known in that case can be utilized. Therefore, the overlaps between different position states no longer need to be avoided. In experiments with trapped ions, one can therefore consider smaller step sizes and thus run the walk for more steps before the required Lamb-Dicke approximation breaks down \cite{Matjeschk2012}.

Secondly, the transformation of the nQW into a QW with orthogonal position states encodes the properties of the non-orthogonality into the initial state of the QW. Therefore, utilizing the overlap, one can prepare several interesting initial states directly, in particular those which are extended over several positions, such as considered in \cite{Valcarcel2010}. In contrast, in the orthogonal case with a localized initial state, the preparation requires a more elaborate process, including several additional operations, some of which must involve a breaking of the translational symmetry. Such a preparation process would severely decrease the fidelity of the experiment.

Furthermore, we show how the initial state can be shifted in momentum space by including an additional operation into the walk operator. This allows for the control of the scaling and the measurement of the dispersion relation, providing a benchmarking tool for the QW. Finally, we use this method of momentum shifts to implement Bloch oscillations \cite{Esaki1970} as an example for the range of experiments with nQWs, which can be readily implemented using state-of-the-art technology.

Throughout the paper, we relate our theory to the trapped-ion setting (Fig. \ref{Fig:zero}). However, it applies to arbitrary unitary nQWs in any dimension.

We consider the Hilbert space $\HH=\ell_2(\Ir)\otimes\Cx^2$, with $\ell_2(\Ir)$ the position space and $\Cx^2$ the coin space. The (normalized but not orthogonal) position states $\ket{\al_x}$, with $x\in\Ir$, form a basis of $\ell_2(\Ir)$. The coin states are given by the $\sigma_z$-eigenstates $\ket{c_+}$ and $\ket{c_-}$, where $\sigma_{x,y,z}$ denote the Pauli matrices. We assume the initial state of the nQW to be localized at the origin, that is, $\rho_0=\ket{\al_0}\bra{\al_0}\otimes \rho_{00}$ with $\rho_{00}=\ket{c_+}\bra{c_+}$ \footnote{Although our method applies to arbitrary initial states, we fix $\rho_{00}=\ket{c_+}\bra{c_+}$ for convenience.}.
One step of the nQW is given by the application of the walk operator $W=S\cdotp \left(\idty\otimes C\right)$, which is composed of a unitary coin operator $C$ and a unitary shift operator $S$. The latter acts as
\beq\label{Eq:shiftop}
S\ket{\alpha_k}\otimes\ket{c_\pm}=\ket{\alpha_{k\pm1}}\otimes\ket{c_\pm}.
\eeq
In fact, starting from $\ket{\alpha_0}$, the subsequent application of $S$ defines all other position states. Hence, their overlap $\braket{\al_k|\al_l}$ is translation invariant. Modeling the trapped-ion systems (Fig. \ref{Fig:zero}), we define the overlap function
\beq\label{EQ:overlap}
g(k)=\braket{\alpha_x|\alpha_{x+k}} =\exp(-k^2/\sigma^2)
\eeq
for all $x$, where $\sigma$ determines the overlap between different position states.

The probability to find the walker at position $\ket{\al_x}$ after $t$ steps is related to the projector $F_x=\ket{\al_x}\bra{\al_x}$ \cite{Schmitz2009a, Matjeschk2012}. That is,
\begin{align}
\label{Eq:probnonorth}
P_{t}(x)&= \frac{ \tr \Bigl( \bigl( F_x\otimes\idty \bigr) \cdotp W^t\,\rho_0\,W^{-t} \Bigr) }{\tr\bigl( G \cdotp\rho_0 \bigr) },
\end{align}
where we introduced $G=\G\otimes\idty$ with the Gram matrix $\G=\sum_x\ket{\al_x}\bra{\al_x}$ for the normalization. Note that the normalization is independent of the step number $t$, because $[S,G]=0$, which can be checked using the unitarity of $S$.

\begin{figure}
\includegraphics[width=8.6cm]{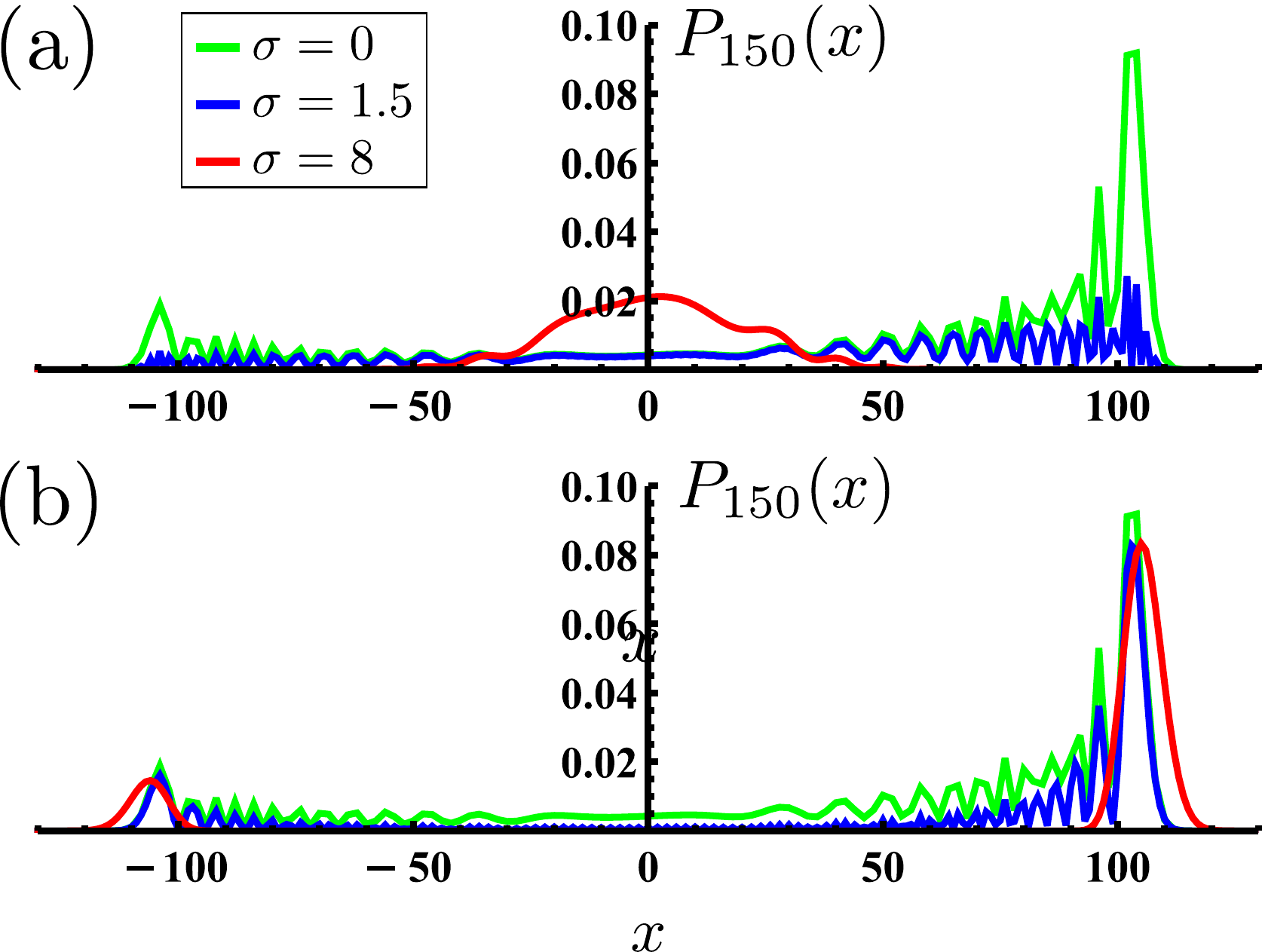}
\caption{(Color online) Position probability distribution $P_t(x)$ \eqref{Eq:probnonorth} after $t=150$ steps of a nQW with overlap function $g(k)$ \eqref{EQ:overlap} and (a) the experimental coin $C_E$ and (b) the Hadamard coin $C_H$ for $\sigma=\{0, 1.5, 8\}$ (green, blue, red). (For the green curve, only points $x$ with $P(x)\neq0$ are connected.) In the orthogonal case (green) the probability distributions of both types of walks are equal, however for large overlaps they differ significantly (blue, red). In case (a), the probability distribution approaches a Gaussian shape centered at the origin of the walk, as $\sigma$ is increased. The spreading, which is still linear in the step number $t$, is vastly reduced \cite{Matjeschk2012}. In the Hadamard case (b), the probability distribution approaches a shape consisting of two Gaussian peaks centered around $\pm t/\sqrt{2}$. Thus, the (linear) spreading is increased, as the probabilities between the peaks vanish. The initial state is $\rho_0=\ket{\al_0}\bra{\al_0}\otimes\ket{c_+}\bra{c_+}$.
}
  \label{Fig:one}
\end{figure}

In Fig.~\ref{Fig:one}, the position probability distributions of nQWs with two different realizations of $C$ are illustrated. The first one (Fig. \ref{Fig:one}a) is with the coin operator $C_E=\exp\left(i\,\pi/4\:\sigma_y\right)$. This type of coin has been implemented experimentally with the initial state $\rho_0$ \cite{Schmitz2009a, Matjeschk2012, Zaehringer2010}. We therefore refer to it as the experimental walk. The second coin operator (Fig. \ref{Fig:one}b) is the Hadamard matrix, which can be written as $C_H=\sigma_z\cdotp C_E$. The coin operators $C_E$ and $C_H$ are very similar and indeed, the probability distributions are equal in the orthogonal case ($\sigma=0$). However, in the case of large overlaps ($\sigma\gtrsim 1$), they show significantly different behaviour.

In the following we will transform the nQW into an orthonormal basis and show that $W$ generates a QW in that basis, but with the initial state being in a superposition of several position states.

The Gram matrix $\G$ gives the relation between the basis $\{\ket{\al_x}\}$ and its dual basis $\{\ket{\al'_x}=\G^{-1}\ket{\al_x}\}$, fulfilling $\braket{\al_x|\al'_y}=\delta_{xy}$ \cite{Daubechies1992}. This allows us to define an orthonormal basis $\{\ket{e_x}=\G^{-1/2}\ket{\al_x}\}$ with $\G^{-1/2}$ being hermitian.
Since the shift operator $S$ commutes with $G$, its action in the orthonormal basis is
\beq\label{Eq:shiftoporthog}
S\ket{e_k}\otimes\ket{c_\pm}=\ket{e_{k\pm1}}\otimes\ket{c_\pm}.
\eeq
That is, $W$ also defines a QW in the orthonormal basis. Since the probability to find the walker in position $\ket{e_x}$ is related to the projector $\G^{-1/2}\,F_x\,\G^{-1/2}$, we can transform Eq. \eqref{Eq:probnonorth} to
\begin{align}
\label{Eq:proporth}
P_{t}(x)&=\tr\Bigl( \bigl(\ket{e_x}\bra{e_x}\otimes\idty\bigr)\cdotp W^t\, \widetilde{\rho}_0 \,W^{-t}\Bigr),
\end{align}
where the initial state amounts to
\begin{align}
\widetilde{\rho}_0
&=\frac{G\widehat{\rho}_0G}{\tr(G\widehat{\rho}_0G)}
\end{align}
with $\widehat{\rho}_0=\ket{e_0}\bra{e_0}\otimes\rho_{00}$. In particular, $\widetilde{\rho}_0$ is extended over several position states according to the overlap function $g(k)$. However, the fact that $\widetilde{\rho}_0$ is extended does not already imply the properties of the nQW. As shown in Fig. \ref{Fig:one}, the experimental and the Hadamard walk show entirely different spreading, although they differ only by the coin operator. In the following we investigate the properties of the nQW using Fourier methods and asymptotic perturbation theory \cite{Amba, Grimmett, Ahlbrecht2011}.

Given a vector $\psi = \sum_{x} \Ket{e_x}\otimes \ket{\psi_x}$ in $\ell_2(\Ir)\otimes\Cx^2$, its Fourier transform in the conjugate momentum space $L^2\bigl([-\pi,\pi),\Cx^2\bigr)$ amounts to $\psi(p)=\sum_{x} e^{i p\cdot x}\ket{\psi_x}$. That is, we consider the Fourier transformed vectors as $\Cx^2$-valued functions of $p$.

\begin{figure}
\includegraphics[width=8.6cm]{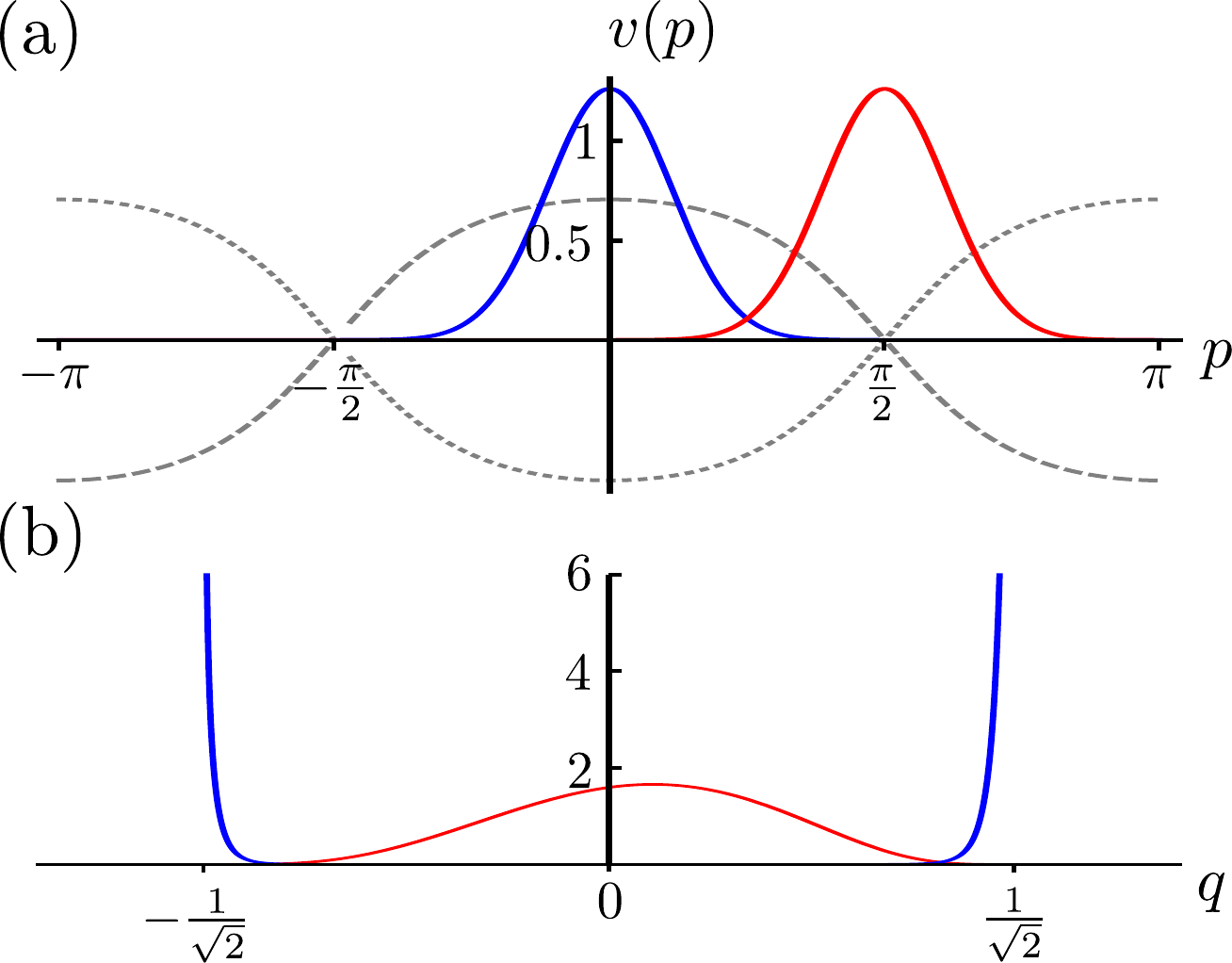}
\caption{(Color online) (a): Group velocities $v_k(p)$ (grey, dotted/dashed) of the Hadamard-walk and $||\widetilde{\rho_0}||(p)/(2\pi)$ of the initial states localized at $p=0$ (blue) and $p=\pi/2$ (red, corresponds to the experimental walk) with $\sigma=4$. (b): Asymptotic position probability distribution $P_\infty(q)$ for each initial state (blue, red). For each initial state only the group velocities around their points of localization determine the position probability distribution. That is, since the blue initial state (a) is localized at $p=0$, where $v_k(0)\approx\pm1/\sqrt{2}$, $P_\infty(q)$ consists of two peaks moving away from the origin with that velocity (b). In contrast, the red initial state (a) is centered around $v_k(\pi/2)\approx0$, which leads to a localized asymptotic position probability distribution $P_\infty(q)$. The coin part of the initial states is $\rho_{00}=\ket{c_+}\bra{c_+}$.
}
  \label{Fig:two}
\end{figure}

The walk operator $W$ is translation invariant on $\ell_2(\Ir)$ and
thus acts as a multiplication operator in momentum space, i.e. $\bigl(W\psi\bigr)(p)=W(p)\psi(p)$, with $W(p)=S(p)\cdotp C$ and $S(p)=\exp(i\,p\,\sigma_z)$. From the eigendecomposition
\beq
W(p)=\sum_{k=1}^{2} e^{i\omega_k(p)} P_k (p),
\eeq
we obtain the dispersion relations $\omega_k(p)$ and the corresponding eigenvectors $\psi_k(p)$, with $P_k (p)$ denoting the projector onto $\psi_k(p)$. The eigenvectors $\psi_k(p)$ define Bloch waves with distinct momentum $p$.
The role of the dispersion relations $\omega_k(p)$ is the same as for a particle in a periodic potential, e.g. an electron in a solid-state system. It encodes the fundamental transport properties of that system. In particular, the group velocities $v_k(p)=\text{d}\omega_k(p)/\text{d}p$ (Fig. \ref{Fig:two}a) determine the spreading behaviour of the initial state of the QW. Precisely, the ballistic order of the spreading (i.e. linear in $t$) can be captured by the time-asymptotic position probability distribution $P_\infty(q)$, where $q\in[-1,1]$ denotes the asymptotic scaled ($\propto 1/t$) position variable. $P_\infty(q)$ can be computed as the inverse Fourier transform of the characteristic function
\beq
C(\lambda)=\int_{[-\pi,\pi)} \text{d}p\:\tr\left(\widetilde{\rho_0}\left(p\right)\cdotp e^{i\lambda V(p)}\right)
\eeq
with $V(p)=\sum_k v_k(p) P_k$ the group-velocity operator \cite{Ahlbrecht2011}. The initial state amounts to $\widetilde{\rho_0}(p)=\lvert g(p)\rvert^2\cdotp\rho_{00}$, where $g(p)$ is the Fourier transform of the overlap function $g(k)$. Therefore, for each momentum $p$, $\lvert g(p)\rvert^2$ determines the influence of the corresponding group velocities $v_k(p)$ to the asymptotic probability distribution.

The group velocities of the cases $C_E$ and $C_H$ are the same, but shifted by $p=\pi/2$ (Fig. \ref{Fig:two}a). Therefore, in the orthogonal case both walks lead to the same probability distribution, since $\vert g(p)\vert$ is constant in $p$. That is, all velocities $v_k(p)$ occur with equal weight, leading to non-zero probabilities in the whole range $x\in\left[-t/\sqrt{2}, t/\sqrt{2}\right]$ (Fig. \ref{Fig:one}). The maximal velocities $v_k(p)=\pm 1/\sqrt{2}$ play a special role by the formation of caustics, leading to the well-known peaks at $x=\pm t/\sqrt{2}$ \cite{Ahlbrecht2011}.

In the non-orthogonal case, $g(p)$ is localized at $p=0$, such that only the related group velocities occur in the nQW (Fig. \ref{Fig:two}a). In the Hadamard walk, $v_k(0)=\pm 1/\sqrt{2}$, i.e. the velocities that are also most pronounced in the orthogonal case, whereas in the experimental case $v_k(0)=0$, such that the position probability remains at the initial position. 

Due to the small but finite width of $g(p)$, also group velocities close to $p=0$ influence the nQW. Since in the case $C_E$ they change strongly around $p=0$, the width of the peak in position space increases linearly in $t$ (See \cite{Matjeschk2012} for numerical results). Similarly, also in the case $C_H$ the widths of the two peaks in position space increase asymptotically linearly in $t$, but at a much smaller rate (Cf. the finite widths of the peaks in fig. \ref{Fig:two}b).

In the following we introduce a method to shift the dispersion relation in momentum space and thus to change the group velocity of the nQW. For a momentum-shift of the amount of $\Theta$, we apply after each step the operator $\idty\otimes R(\Theta)=\idty\otimes \exp(i\,\Theta\,\sigma_z)$. Due to the identity $R(\Theta)S(p)=S(p+\Theta)$, this is equivalent to a nQW with the effective walk operator
\begin{align}
W_\Theta (p)&=S(p+\Theta)\cdotp C\, .
\end{align}
The time evolution is then determined by the group velocities $v_k(p+\Theta)$. Thus, using the experimental coin $C_E$, it is possible to achieve the spreading of a Hadamard walk by including the operator $R(-\pi/2)$ into $W$. In fact, since $C_H=\sigma_zC_E$, the momentum shift with $\Theta=-\pi/2$ compensates for the $\sigma_z$-factor, up to a complex phase.

The momentum-shift method allows for the experimental determination of the dispersion relations $\omega_k(p)$ by implementing the nQW with $W_\Theta$ for several values of $\Theta\in[-\pi,\pi]$ and determining the scaling of the position probability distribution for each. This is particularly important if the walk operator is not exactly known, e.g. due to experimental imperfections and in the regime of a high number of steps.

\begin{figure}
\includegraphics[width=8.6cm]{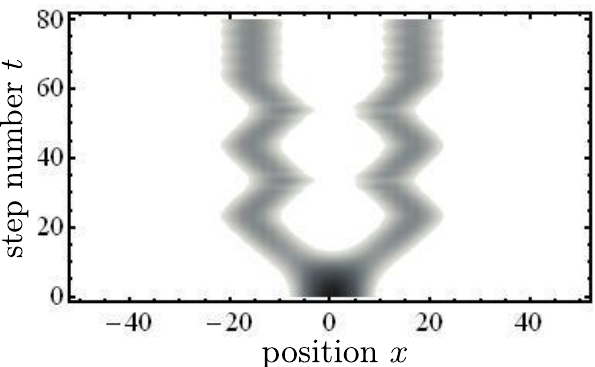}
\caption{Probability density (greyscale with black: $P_t(x)=1$) of a Hadamard-nQW ($\sigma=14$) with Bloch oscillations. After 20 steps of the nQW (Cf. fig. \ref{Fig:one}b), the Bloch oscillations are switched on with $\Delta\Theta=\pi/10$, such that the positions of the peaks oscillate with a period of 20 steps and an amplitude of 5 positions. The Bloch oscillations are switched off after 65 steps, a point where the group velocity is zero (Fig. \ref{Fig:two}a), such that the peaks remain at their position during the remaining nQW. The expectation values of the harmonic-oscillator occupation-number operator $N$ range during the oscillations from $\braket{N_{min}}=1.3$ to $\braket{N_{max}}=2.7$, which is detectable with state-of-the-art trapped-ion technology \cite{Wineland1998}.
}
  \label{Fig:three}
\end{figure}

In semiconductor-superlattices, the driving force for Bloch oscillations is implemented by a static external electric field, leading to a linear drift of the electrons' momentum. The periodic band structure causes the oscillatory behaviour of the electrons, detectable by optical methods \cite{Leo1992}. The analog of this motion can be implemented in QWs by applying the momentum-shift operator $R(t\cdotp \Delta\Theta)$ (modulo $2\pi$) at the $t$-th step (for every $t$), which implies a shift in momentum space by $\Delta\Theta$ at each step of the nQW. The walker thus experiences different group velocities at each step of the walk (Fig. \ref{Fig:two}a), which results in an oscillating behaviour in position space, in contrast to linear spreading with a constant group velocity (Fig. \ref{Fig:three}). Note, that this effect does not require the non-orthogonality of the position states. However, the simple shape of the position probability distribution of a nQW (two distinct Gaussian peaks) can reduce the effort for detection.

A convenient system for the experimental implementation are trapped ions. Recently, one-dimensional QWs with three, resp. 23 steps have been realized in the phase space of the harmonic motion \cite{Schmitz2009a, Zaehringer2010} and a protocol for the extension to 100 steps has been proposed \cite{Matjeschk2012}. The number of steps in these experiments was limited by two requirements: On the one hand, the motional amplitude of the ion needed to remain small, because the implemented protocols \cite{Travaglione2002} were designed assuming the Lamb--Dicke approximation \cite{Wineland1998}. On the other hand, the step size $\Delta\alpha$ (Fig. \ref{Fig:zero}) was chosen sufficiently large in order to minimize the overlap between neighbouring position states.
For the implementation of Bloch oscillations, a small step size is favoured, leading to a significantly higher possible number of steps. A possible choice of parameter values for a Lamb--Dicke parameter of $\eta\leq0.3$, as in Ref. \cite{Matjeschk2012}, is given in Fig. \ref{Fig:three}. The momentum-shift operator $R(\Theta)$ can be implemented by shifting the phase of the driving light fields with respect to the relative phase of the coin states \cite{Hanneke2010}. The positions of the peaks can be determined using state-of-the-art blue-sideband protocols \cite{Wineland1998}.
During preparation of our manuscript we were made aware of a related implementation of Bloch oscillations \cite{Silberhorn2011}. In contrast to that approach, our method does not require a position-dependent coin operator and may therefore require less technical effort.

In summary, the transformation of nQWs into orthogonal ones allows for an intuitive understanding of their properties in terms of the dispersion relations. Hence, by the momentum-shift method it is possible to change the spreading behaviour, to determine the dispersion relation, and to implement - by the correspondence to solid-state systems - the analog effect of Bloch oscillations. Therefore, the non-orthogonality can be exploited and need not to be avoided, leading to a higher number of steps and a new range of experiments with available technology. Furthermore, nQWs can be considered for modeling transport processes in complex systems and may lead to a better description than previous approaches.

\paragraph{Acknowledgements}
This work was supported by the EU-project COQUIT (grant number: 233747) and the DFG (Forschergruppe 635). We also thank David Schwandt for discussions.

%

\end{document}